# Tunneling transport in a few monolayer-thick $WS_2$/graphene heterojunction

Takehiro Yamaguchi[1], Rai Moriya[1,a)], Yoshihisa Inoue[1], Sei Morikawa[1], Satoru Masubuchi[1,2], Kenji Watanabe[3], Takashi Taniguchi[3], and Tomoki Machida[1,2,b)]

[1]*Institute of Industrial Science, University of Tokyo, 4-6-1 Komaba, Meguro, Tokyo 153-8505, Japan*

[2]*Institute for Nano Quantum Information Electronics, University of Tokyo, 4-6-1 Komaba, Meguro, Tokyo 153-8505, Japan*

[3]*National Institute for Materials Science, 1-1 Namiki, Tsukuba 305-0044, Japan*

This paper demonstrates the high-quality tunnel barrier characteristics and layer number controlled tunnel resistance of a transition metal dichalcogenide (TMD) measuring just a few monolayers in thickness. Investigation of vertical transport in $WS_2$ and $MoS_2$ TMDs in graphene/TMD/metal heterostructures revealed that $WS_2$ exhibits tunnel barrier characteristics when its thickness is between 2 to 5 monolayers, whereas $MoS_2$ experiences a transition from tunneling to thermionic emission transport with increasing thickness within the same range. Tunnel resistance in a graphene/$WS_2$/metal heterostructure therefore increases exponentially with the number of $WS_2$ layers, revealing the tunnel barrier height of $WS_2$ to be 0.37 eV.

---

a)E-mail: moriyar@iis.u-tokyo.ac.jp
b)E-mail: tmachida@iis.u-tokyo.ac.jp

The van der Waals heterostructure of graphene and other layered crystals has recently received considerable attention [1,2] owing to the fact that their weak interlayer coupling allows for the mechanical exfoliation down to a single monolayer. Thus, it is possible to fabricate heterostructures based on different crystal layers connected through van der Waals force, which introduces another degree of freedom and has the potential to create applications in electronics. As a building block for these van der Waals heterostructures, transition metal dichalcogenides (TMDs) such as $MoS_2$ and $WS_2$ have proven particularly popular [3], as unlike graphene they exhibit semiconducting properties. Indeed, high-performance transistors, photovoltaic cells and flexible electronics based on vertically-stacked graphene/TMD heterostructures have already been demonstrated [4,5,6,7,8,9]; combining TMDs with other layered materials has the potential to achieve various functional devices. So far, most studies into these heterostructures have focused on vertical conduction across a graphene/TMD heterointerface through a thermally activated process, in which modulation of the Schottky barrier height gives rise to a large current modulation in graphene/thick-$MoS_2$/metal and graphene/$WS_2$/graphene vertical transistors [4,6,9]. Studies into the transport through thin layers (3 monolayer or less) of TMD, on the other hand, have been quite limited. In thin layer TMDs, one would expect tunneling to be the dominant conduction process, and in fact a highly spin-polarized tunneling has been theoretically predicted in a monolayer $MoS_2$ tunnel barrier [10]. The use of inter-band tunneling through a TMD layer has also been considered as a possible candidate for high performance tunnel field-effect transistor applications [11], thus creating a demand for more detailed studies into the tunneling properties of graphene/TMD heterojunctions. The potential of such layered materials has already been demonstrated by the well-known insulating crystal h-BN, in which the absence of dangling bond on its surface gives rise to high-quality tunnel barrier characteristics down to a thickness of a single atomic layer [12,13,14,15].

In this letter, we look at the feasibility of achieving vertical transport through 2–5 monolayer-thick TMDs of $WS_2$ and $MoS_2$ by means of a graphene/TMD/metal vertical transistor structure with a view to developing a van der Waals heterostructure suitable for use in future electronic applications.

A schematic illustration of the device structure used is given in Fig. 1(a), in which graphene produced by the mechanical exfoliation of Kish graphite was used to fabricate a bottom electrode on a $SiO_2$ 300 nm/$n$-Si(001) substrate. Onto this was fabricated crystalline TMDs of a few monolayers thickness through a combination of the mechanical exfoliation of bulk $WS_2$ or $MoS_2$ crystal (2D semiconductors Inc.) and a dry transfer technique [16,17]; a method that allowed a van der Waals heterojunction to be formed between the freshly cleaved surfaces of the graphene and TMD. Finally, electron beam (EB) lithography (Elionix EBL7500) and EB evaporation was used to produce Au 30 nm/Ti 50 nm or $Ni_{81}Fe_{19}$ 40 nm electrodes. The junction area of the series of devices produced was 1–3 μm$^2$. The carrier concentration of the graphene layer was controlled by applying a gate voltage ($V_G$) to the highly doped $n$-Si substrate, with all devices being characterized in a variable temperature cryostat within a temperature range of 4 to 300 K.

Vertical transport across the TMD layer was measured using a three-terminal method, wherein a current bias $I$ was first applied between the contacts C1 and C2 shown in Fig. 1(a), and the voltage difference $V_B$ between contacts C1 and C3 was measured with a voltmeter. Using this method, the contribution of the graphene's series resistance can be eliminated and the resistance of the graphene/TMD/metal junctions is precisely determined. For devices with a junction resistance higher than 1 MΩ, a standard two-terminal $I$-$V$ curve measurement was used, in which such a voltage bias $V_B$ was applied between contacts C1 and C2 and the resulting current flow through the device was measured.

The $I$-$V$ curves obtained from the graphene/2-monolayer $WS_2$ and $MoS_2$/metal vertical structures are shown in Fig. 1(b), with the $I$-$V$ curve for a few-layer graphene/2-monolayer h-BN/metal structure also shown for comparison. All $I$-$V$ curves were measured at 300 K, and $V_G$ was adjusted to the Dirac point $V_{DP}$ of the graphene layer. The current values are normalized with its junction area. This clearly shows that the 2-monolayer $WS_2$, $MoS_2$, and h-BN layers all exhibit non-linear $I$-$V$ characteristics. From this, the zero-bias conductance $G = dI/dV_B$ at $V_B = 0$ V was extracted, which is plotted as a function of temperature in Fig. 1(c). This demonstrates that $G$ exhibits only a very small temperature dependence in all of the layered materials, which combined with the non-linear nature of the $I$-$V$ curves, suggests that vertical transport through these devices is by tunneling. That the h-BN displays almost no temperature dependence of $G$ has been previously observed, and suggests a large tunneling barrier height of 1.5–3 eV [12,13,14]. Both 2-monolayer $WS_2$ and $MoS_2$ have a similarly small temperature dependence, although the change in $G$ between 10 and 300 K in these devices is notably more pronounced than in h-BN. Moreover, the larger zero bias conductance $G$ of $MoS_2$ and $WS_2$ indicates that the tunnel barrier height of these materials is lower than that of h-BN. Using this data, the tunnel barrier height can be roughly estimated by considering the thermal smearing of the direct elastic tunneling contribution [18,19]. For this, the temperature dependence of the zero-bias conductance $G(T)$ data is fitted using the function:

$$G(T) = G_0 CT / \sin CT,$$

where $G_0$ is the zero-bias conductance at 10 K, $C = 2\pi^2 k_B d \sqrt{2m^*} / \sqrt{h^2 \varphi}$, $k_B$ is the Boltzmann constant, $d$ the thickness of the tunnel barrier, $m^*$ the effective mass inside the tunnel barier, $h$ the Plank constant, and $\varphi$ the tunnel barrier height. For the TMDs, a monolayer thickness of 0.65 nm and tunneling effective mass $m^* = 0.6 m_0$ ($m_0$ denotes electron mass) for the conduction band electron was used for both $MoS_2$ and $WS_2$ [20,21]. The results of this fitting are plotted as a dashed line in Fig. 1(c), which demonstrates reasonably good agreement between the experimental and fitted data in terms of a tunnel barrier height of $\varphi$ ~0.41 and 0.45 eV for $MoS_2$ and $WS_2$, respectively. These results suggest the tunnel barrier height is lower in these TMD layers than in h-BN, which is discussed in more detail later in relation to its layer number dependence. Nevertheless, this small dependence on temperature is quite distinct from previous observations of graphene/thick-$MoS_2$/metal vertical field effect transistors with four or more $MoS_2$ monolayers [4,9], in which the presence of a Schottky barrier at the graphene/$MoS_2$ interface gave rise to strong asymmetry in the *I-V* curve and a significant dependence on temperature with regards to conductance. Clearly, the mechanism by which conductance is achieved through 2 monolayer-thick $MoS_2$ and $WS_2$ is quite different to that encountered in a much thicker $MoS_2$ layer, and thus we believe that by fabricating an extremely thin TMD layer tunnel conductance becomes dominant over thermionic emission.

The change in junction conductance $G$ in both $WS_2$- and $MoS_2$-based vertical heterostructures in response to a sweep of the gate voltage $V_G$ applied to the $n$-Si substrate at 300 K is shown in Figs. 2(a) and (b), in which the horizontal axis is plotted as the difference from the Dirac point of the graphene layer $V_{DP}$. It is apparent from this that the conductance of the two monolayer thick $WS_2$-based device ($N = 2$) experiences what is essentially a symmetrical change with respect to $V_G$. A similar dependence of conductance on gate-voltage has been previously observed in graphene/h-BN/graphene tunnel transistors [14], and can be explained by the gate dependence of tunneling conductance through the device being dominated by the graphene's density of state rather than the change in tunnel barrier height. This behavior remained unchanged with a $WS_2$-based device in which $N = 4$, as shown in Fig. 2(a).

In the case of $MoS_2$, the dependence of conductance on $V_G$ was found to be very similar to $WS_2$ when $N = 2$, but becomes significantly asymmetric when $N = 4$, meaning that the conductance monotonically increases with $V_G$. Such behavior is reminiscent of thermionic emission across the Schottky barrier that has been observed at the graphene/$MoS_2$ interface in a graphene/thick-$MoS_2$/metal vertical field effect transistor [4,9], wherein the modulation of Schottky barrier height with gate voltage has a more pronounced effect than the change in graphene's density of state. The asymmetry of the $V_G$ dependence was defined as $\eta = G(V_G = V_{DP} +26 \text{ V})/G(V_G = V_{DP} -26 \text{ V})$, the results of which are plotted in Fig. 2(c). The fact that $\eta$ remains essentially constant for $WS_2$ within a thickness range of $N = 2$–5 suggests that its vertical transport mechanism does not change, whereas the significant increase in $\eta$ evident in $MoS_2$ when $N = 4$ indicates that thermionic emission becomes the dominant transport mechanism at this thickness.

Variation in the transport behavior with layer number is also evident in the way that $G$ changes with respect to temperature, as shown in Figs. 2(d) and (e). Note that although $V_G$ is adjusted to the Dirac point of the graphene layer for comparison, gate voltage does not actually affect these particular results. Thus, there is a very weak correlation between temperature and $N$ in the case of $WS_2$, with Fig. 2(d) showing that $G(T)$ is very similar at $N = 2$ and 5. In contrast, as shown in Fig. 2(e), the $G(T)$ of $MoS_2$ changes significantly between $N = 2$ and 5, and increases with temperature at $N = 5$. This much greater temperature dependence provides further evidence that thermionic emission becomes dominant in $MoS_2$ when it is thicker than $N = 3$. Qualitatively, the thermionic emission current across a graphene/TMD interface is weakly dependent on the thickness of the TMD layer, whereas the tunneling conductance exponentially decreases with TMD thickness. Thus, with a low barrier heterojunction, a transition in conduction mechanism from tunneling to thermionic emission is expected. The results of this study therefore imply that the band offset of graphene/$MoS_2$ is lower than that of graphene/$WS_2$, which is certainly consistent with recent calculations [22,23]. On the basis of this, a graphene/$WS_2$ heterojunction is considered to be more suitable for use as a tunnel barrier than graphene/$MoS_2$.

The tunnel barrier height of the $WS_2$ layer was evaluated from the thickness dependence of the zero bias resistance area products $RA = 1/G$. The low bias $I$-$V$ curves measured at 300 K for different $WS_2$ thickness are plotted in Fig. 3(a), wherein the gate voltage $V_G$ is again adjusted to the Dirac point of the graphene layer. It is evident from this that the $I$-$V$ curve changes systematically with the number of $WS_2$ layers, and so the $RA$ obtained from the $I$-$V$ curves are plotted in Fig. 3(b) with respect to the layer thickness *t*. For comparison, the *R*A for a h-BN tunnel barrier where $N$ = 1–4 is also shown. The thickness of both materials was calculated on the assumption that the monolayer thickness of $WS_2$ and h-BN is 0.65 and 0.34 nm, respectively. In both materials, the junction resistance can be seen to be weakly dependent on the electrode material. Furthermore, in the case of h-BN, *R*A increases exponentially with *N* and is therefore consistent with previously observed results[12,13,15]. A similar exponential increase in *R*A with respect to thickness is also evident in the case of the $WS_2$ tunnel barrier. In an ideal tunnel barrier, the junction resistance should follow the relationship:

$$\log(RA) \propto \left(4\pi\sqrt{2m^*\varphi}/h\right)t,$$

where $m^*$ is the effective mass inside the tunnel barrier, and $\varphi$ the average tunnel barrier height. The fact that a linear relationship can be observed between $\log(RA)$ and *t* in Fig. 3(b) therefore indicates that a high quality tunnel barrier has been achieved with $WS_2$, without any pinholes. Moreover, given the fact that the slope of $\log(RA)$ vs. *t* is smaller with $WS_2$ than h-BN despite using the same electrodes provides direct evidence that $WS_2$ has a smaller barrier height; an average barrier height of 3.0 and 0.37 eV being obtained for h-BN and $WS_2$, respectively, with an effective mass of $m^* = 0.5m_0$ and $0.6m_0$. Significantly, this barrier height of h-BN is within the range of previous reports suggesting it is between 1.5 to 3.0 eV [12,13,14]. The tunnel barrier height in Gr/TMD/metal heterostructure is determined by the band offset at graphene/TMD interface ($\Phi_{GT}$) and metal/TMD interface ($\Phi_{MT}$) [24]. If the Schottky-Mott rule is assumed to hold, then these band offsets can be expressed as $\Phi = \phi - \chi$, where $\phi$ denote work function of the metal or graphene, $\chi$ the electron affinity of the TMD [25]. Recent density functional theory calculations have shown the electron affinity of $WS_2$ to be $\chi \sim$ 4.0 eV [22]; and so considering the fact that the work function of graphene, Ti, and Ni$_{81}$Fe$_{19}$ are $\phi$ = 4.6, 4.3, and 5.0 eV, respectively [26,27], we calculated $\Phi_{GT}$ = 0.6 eV and $\Phi_{MT}$ = 0.3 eV for Ti and 0.7 eV for Ni$_{81}$Fe$_{19}$, respectively. The tunnel barrier height in the graphene/WS$_2$/metal heterostructure is therefore expected to be $\varphi = (\Phi_{GT} + \Phi_{MT})/2$ = 0.45 and 0.65 eV, respectively. These values are reasonably close to the tunnel barrier height that was determined experimentally for our devices. The discrepancy between calculation and experiment can be attributed to the decrease in actual $\Phi_{MT}$ that is caused by the pinning of metal's Fermi level at metal/MoS$_2$ interface [28]: this contribution also makes band offset to be insensitive to the metal work function as we observed in our results.

In conclusion, the results obtained in this study clearly demonstrate that WS$_2$ can be used as a high-quality, layered tunnel barrier material, and that its thickness can be effectively controlled at monolayer-thick scale. Meanwhile, the lower tunnel barrier height and greater spin orbit coupling of WS$_2$ compared to h-BN means that this could have important implications for the development of electronic and spintronic devices.

**Acknowledgements**

This work was partly supported by a Grant-in-Aid for Scientific Research into the Science of Atomic Layers from the Japan Society for the Promotion of Science (JSPS), and a Grant-in-Aid for Scientific Research into Innovative Areas and Project for Developing Innovation Systems from the Ministry of Education, Culture, Sports and Technology (MEXT). S. Morikawa acknowledges the JSPS Research Fellowship for Young Scientists.

**Figure captions**

Figure 1

(a) Schematic illustration of 2D vertical heterostructures. (b) Current-voltage (*I-V*) characteristics of graphene/MoS$_2$, WS$_2$ and h-BN/metal devices at 300 K. The number of MoS$_2$, WS$_2$, and h-BN layers was 2 and $V_G$ was set to the Dirac point of the graphene electrode. (c) Temperature dependence of zero bias conductance *G* for graphene/MoS$_2$, WS$_2$ and h-BN/metal devices. Dashed lines indicate the fitted results based on the thermal smearing contribution of tunneling conductance.

Figure 2

(a,b) Relationship between *G* and ($V_G$–$V_{DP}$) for 2 and 4 layers of (a) WS$_2$ and (b) MoS$_2$ measured at 300 K. (c) Change in asymmetry $\eta$ with respect to the number of layers of WS$_2$ or MoS$_2$, *N*. Dashed line indicates $\eta$ =1. (d,e) Temperature dependence of junction conductance *G*(*T*) normalized to its value at 10 K for (d) WS$_2$ and (e) MoS$_2$.

Figure 3

(a) Low bias *I-V* characteristics at room temperature of a graphene/WS$_2$/Ni$_{81}$Fe$_{19}$ structure with a varying number of WS$_2$ layers. $V_G$ was adjusted to the $V_{DP}$ of the graphene layer. (b) Thickness dependence of zero bias resistance area product *RA* for h-BN and WS$_2$. Solid circles and solid squares represent devices using Ni$_{81}$Fe$_{19}$ and Ti electrodes, respectively.

Figure 1

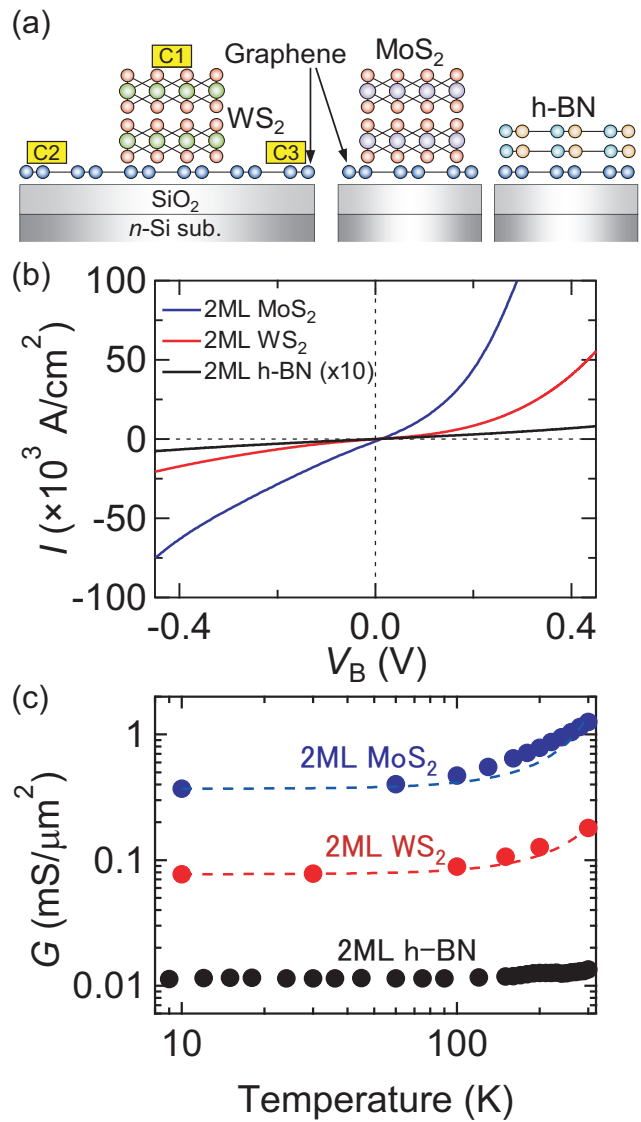

Figure 2

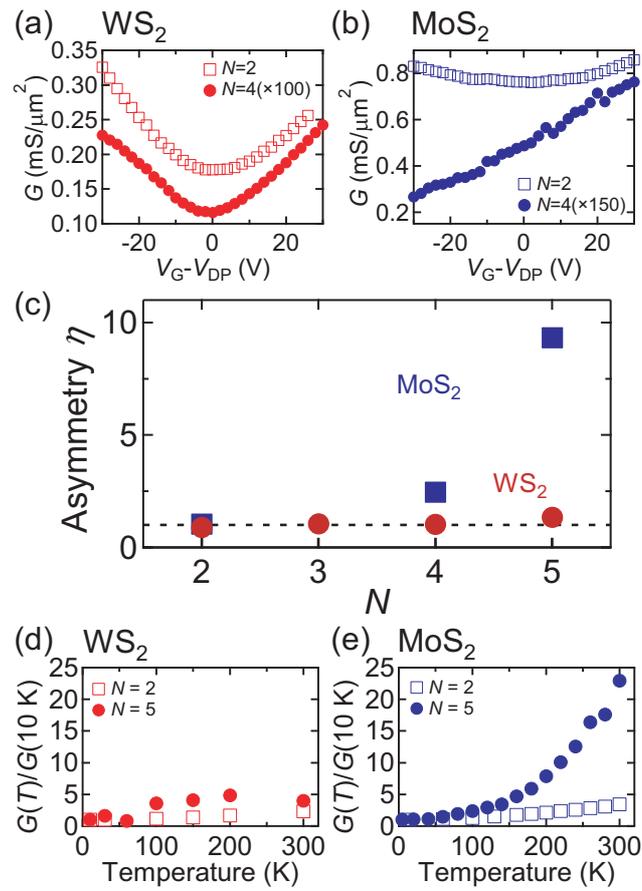

Figure 3

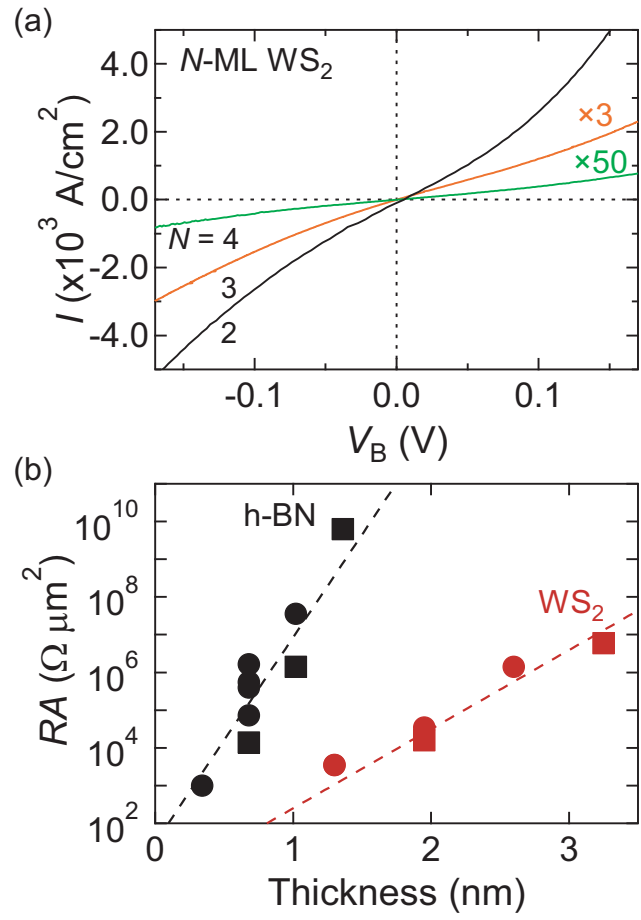